\documentclass[twocolum,5p]{elsarticle}

\usepackage{amssymb}
\usepackage{amsmath}
\usepackage{graphicx}
\usepackage[colorlinks,hyperindex]{hyperref}
\usepackage[utf8x]{inputenc}
\usepackage{float}
\usepackage{braket}

\journal{Physica B.}

\begin{document}

\begin{frontmatter}

\title{Effective quasiparticle approach for a Cavity-QDots System}
\author[un1]{F. G\'omez}

\author[un2]{J. P.  Restrepo Cuartas}
\author[un1]{B. A.  Rodr\'iguez Rey}
\author[un2]{H. Vinck-Posada\corref{cor1}}
 \address[un1]{Grupo de Física Atómica y Molecular, Instituto de Física\\Universidad de Antioquia UdeA, Calle 70 No. 52-21, Medellín, Colombia}
 \ead{hvinckp@unal.edu.co}
\address[un2]{Universidad Nacional de Colombia - Sede Bogotá, Facultad de Ciencias, Departamento de Física\\
 Carrera 45 No. 26-85, C.P. 111321, Bogotá, Colombia}

\cortext[cor1]{Corresponding author}
\date{\today}

\begin{abstract}
In this work, we present a quasiparticle strategy to study  the Hamiltonian description of the stationary states for two quantum dots--cavity system.  We consider three different effective schemes of quasiparticles that give an in-depth insight into the physics involved in the Hamiltonian eigenstates for parameters that optimize or minimize the energy gap condition. We analyze features of quantum measures like fractional composition, linear entropy,  and concurrence to observe which one description gives the complete physical information. Our findings show  that  a polaritonic—light-matter quasiparticle—approach catch better the physics contained in the whole regimes considered.
\end{abstract}


\end{frontmatter}


\section{Introduction}

The emission of light in many solid state systems has been of broad interest in the last years~\cite{sanvitto2016road,heindel2017bright,jayaprakash2017ultra}, especially active media like quantum wells~\cite{lagoudakis2004coexistence,christmann2008room,kira1997quantum}, quantum wires~\cite{byelobrov2010low,atlasov2009photonic,arnardottir2017hyperbolic}, and quantum dots (QDs)~\cite{gies2007semiconductor,kreinberg2017emission} embedded in a microcavity. Some of these works show that it is possible to find different ways to get coherent light from the system.  In particular, J. Bloch’s group has shown that it is possible to achieve coherent emission of light with and without population inversion~\cite{deng2003polariton,Azzini11}, which have suggested reviewing the role of quantum correlations between radiation and matter.  Additionally, in the linear quantum regime, i.e., low external pumping, new single photon sources have been developed; they are promising candidates to be included in outstanding quantum technologies~\cite{eisaman2011invited,somaschi2016near,ripka2018room}.
Quantum coherence in QDs molecules—that can be built up isolated or immersed into microcavities—has been accounted for and measured extensively~\cite{chow2016nonlocal,khoshnegar2014toward,tian2015cavity,stanislav,evans2018photon,economou2010optically,calic2017deterministic,heo2017implementation}. A plethora of quantum effects due to quantum correlations in strongly coupled QDs allows the semi-conductor community to develop applications on quantum computing, quantum cryptography, optical bistability (non-linear response)~\cite{tian2015tunneling,BO}, quantum teleportation, entanglement in QD mo\-le\-cu\-les \cite{gauger2008robust}, and induced transparency~\cite{transparenci}.

Theoretically, models that imply the concept of quasiparticles like plasmons, polarons, phonons, excitons, and polaritons, among others, have been widely used to describe condensed matter systems~\cite{Azzini11,electr-tunneling,dipolariton,moleculetunneling,tunneling-e-h}. However, the remaining question is: what is the most useful representation, in terms of quasiparticles, to understand the effective mechanisms that are involved in the physical description of the coupled quantum radiation-matter problem?
To give some insights, that allow understanding the main components of this problem, we tackle a particular situation: two interacting QDs embedded in a microcavity. For this system, we consider three possible effective representations in terms of quasiparticles, which are, non-interacting picture, molecular picture, and polaritonic picture. We focus our analysis on specific situations that show different critical regimes for energy bandgaps.

The paper is organized as follows: in section 2, we describe the system and the theoretical framework, emphasizing different approaches that account for the main features of the quasiparticle scheme. Then, in section 3, we characterize three stationary operational regimes for the system, and we compare the three different quasiparticle schemes to analyze the involved physical mechanisms. Finally, in section 4, we summarize and conclude.

\section{Physical System and Theoretical Framework}

\begin{figure}[tbh]
\centering
\includegraphics[width=0.45\textwidth]{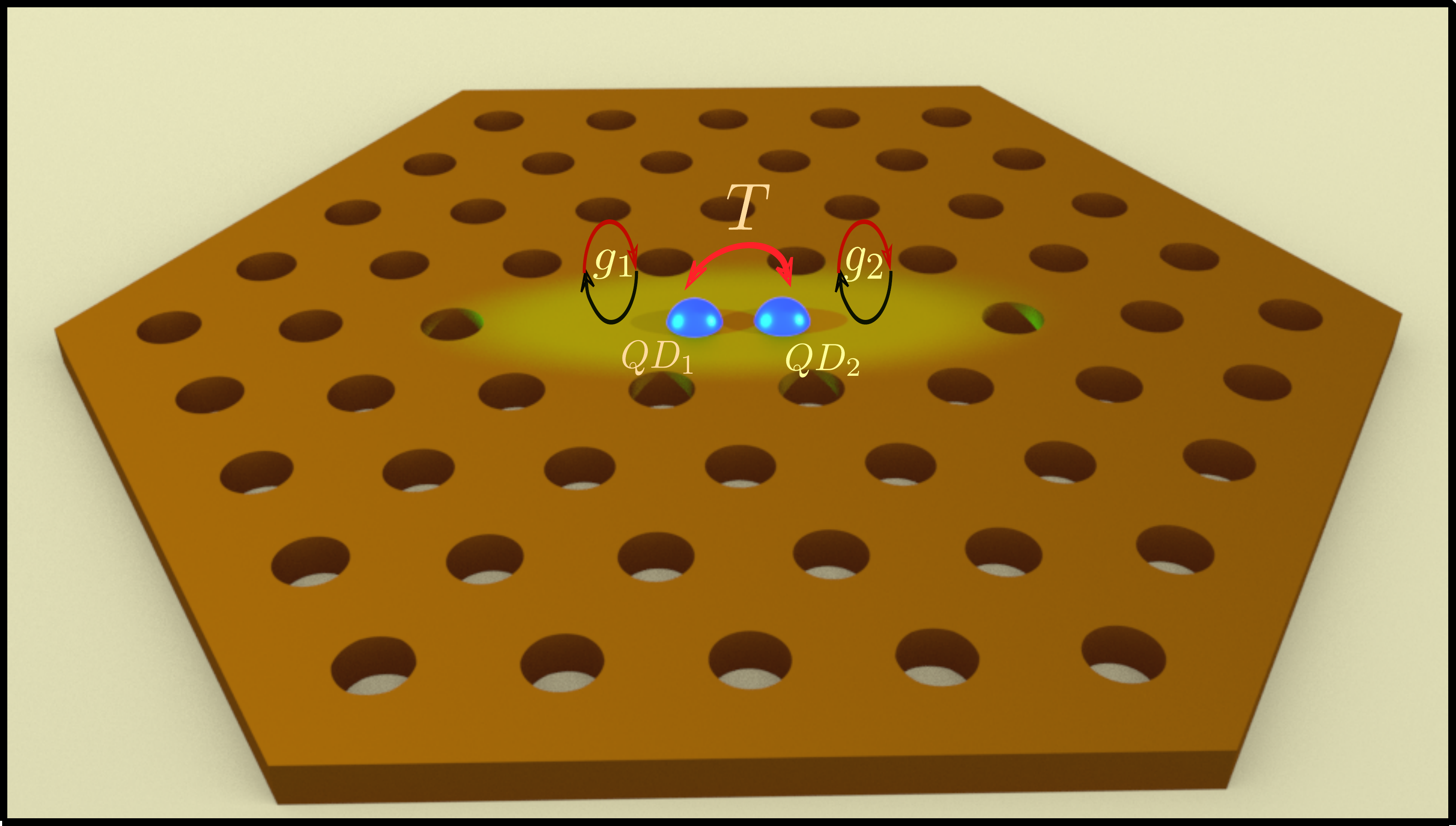}
\caption{Schematic representation of  two interacting quantum dots embedded in a microcavity, $g_1$ and $g_2$ are the interaction radiation-matter couplings and $T$ is the tunnelling coupling constant.}
\label{fig:system}
\end{figure}

We consider two QDs embedded in a microcavity. Each QD can interact with the other and with a single cavity mode $\omega_c$, as depicted in Fig. \ref{fig:system}. The following Hamiltonian models the system:
\begin{align}\label{eq:ham}
H&=\hbar\omega_{1}\sigma^{\dagger}_{1}\sigma_{1} + \hbar\omega_{2}\sigma^{\dagger}_{2}\sigma_{2} +
 \hbar\omega_c a^{\dagger} a
 \nonumber
\\ &+ \hbar g_{1}\left( a \sigma^{\dagger}_{1} +  a^{\dagger}\sigma_{1}\right)  + \hbar g_{2}\left(a\sigma^{\dagger}_{2} +  a^{\dagger}\sigma_{2}\right) \\ &+ \hbar T\left(\sigma_{1}\sigma^{\dagger}_{2} +  \sigma^{\dagger}_{1}\sigma_{2}\right), \nonumber
\end{align}
 where $\omega_{c}$ is the cavity frequency, $\omega_{i}$ (i=1,2) are the QD free frequencies, $g_{i}$ are the light-matter interaction strength coefficients of the QDs with the cavity mode, and $T$ is the tunnelling constant between the two QDs. In the following, we consider $\hbar=1$. As usual, the processes of creation and annihilation of photons in the cavity mode are described by the operators $a^{\dagger}$ and $a$,  while the creation and annihilation of excitations in the QDs are described by the operators $\sigma_{i}^{\dagger}$ ($\sigma_{i}$), respectively.

\begin{figure*}[tbh]
\centering
\includegraphics[width=1.0\textwidth,trim={0 3.5cm 0 3cm},clip]{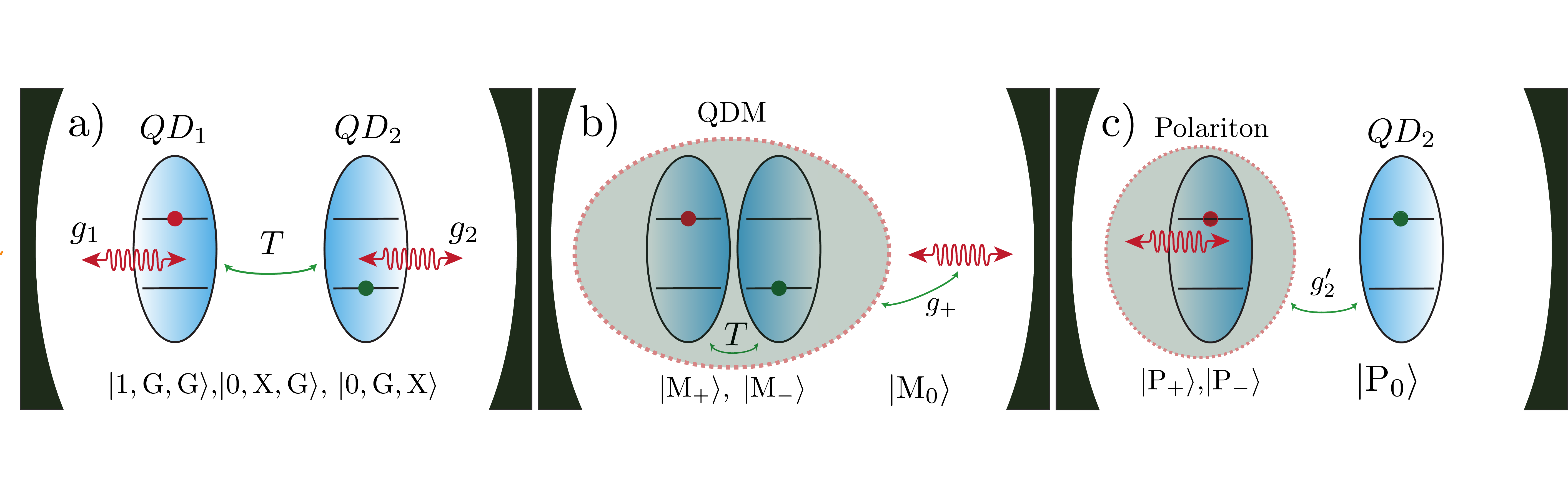}
\vspace{-0.4cm}
\caption{Schematic quasiparticle representations. (a) Two interacting quantum dots and a photonic cavity mode. (b) Effective molecule of quantum dots interacting with a photonic cavity mode and (c)   Effective polariton interacting with a quantum dot.}
\label{fig:quasiparticle}
\end{figure*}

Since three parts compose our system, we discuss three different approaches to assess stationary eigenstates. We take into account an effective quasiparticle picture. In the first case, we use a bare basis, i.e., the basis that corresponds to the eigenvectors of the non-interacting Hamiltonian ---see Fig.~\ref{fig:quasiparticle} (a)---.  Then, in the second case, we discuss the interaction between the cavity mode and the molecular modes that arise from the tunnelling coupling. To do that,  we diagonalize the subspace that involves the isolated matter states getting a new basis which corresponds to molecular dressed states ---Fig.~\ref{fig:quasiparticle} (b)---. Finally, we consider a polaritonic basis; this one is obtained by diagonalization of the subspace composed by one QD state and the photonic mode ---Fig.~\ref{fig:quasiparticle} (c)---.

\subsection{Quasiparticle Pictures.}
To build up the basis sets, previously described,  we constrain the Hamiltonian representation up to the first excitation manifold $(\Lambda_{1})$. The three basis sets correspond to the following cases. 

\subsubsection{ Bare States Picture}

Henceforth, we use the common occupation number notation,  $\ket{\textit{Cavity},\textrm{QD}_1,\textrm{QD}_2}$, for naming the Hilbert space. Then, the  bare basis is shaped like $\ket{1,\textrm{G},\textrm{G}}$, $\ket{0,\textrm{X},\textrm{G}}$,  $\ket{0,\textrm{G}, \textrm{X}}$. The following expression corresponds, in the bare basis, to the matrix representation of the Hamiltonian (equation (\ref{eq:ham})): 

\begin{equation}\label{eq:Hami2}
H=\left (\begin{array}{ccc}
\omega_{c}  & g_{1} & g_{2} \\ 
g_{1} & \omega_{1} & T \\ 
g_{2} & T & \omega_{2}
\end{array}\right)
\end{equation}

\subsubsection{ Molecular Basis Picture}
To take into account the effects of tunnelling interaction, we build up a new basis  using the molecular dressed states of the following Hamiltonian
\begin{align}\label{eq:mono} 
H_{QD_1-QD_2}& =\begin{pmatrix} 
\omega_1 & T \\
T &\omega_2
\end{pmatrix},
\end{align}
the dressed eigenstates are 
\begin{align}\label{eq:autoest}
\ket{ \tilde{m}_+ }&= \cos\theta_m \ket{\textrm{X},\textrm{G}} + \sin\theta_m \ket{\textrm{G},\textrm{X}}; \\
\ket{ \tilde{m}_-} &= -\sin\theta_m \ket{\textrm{X},\textrm{G}} + \cos\theta_m \ket{\textrm{G},\textrm{X}},
\end{align}
and the corresponding eigenvalues are
\begin{equation}
\textrm{E}_{\pm} = \dfrac{1}{2} (\omega_{1}+\omega_{2}) \pm \dfrac{1}{2} \sqrt{(\omega_1-\omega_2)^2+4\textrm{T}^2}.
\label{eq:aval}
\end{equation}
Additionally, we define the ground state for the molecule as $\ket{\tilde{m}_0}=\ket{\textrm{G},\textrm{G}}$. Using these eigenstates along with the bare states of light we construct a new basis defined as: $\ket{M_0}=\ket{1}\otimes \ket{\tilde{m}_0}$, $\ket{M_+}=\ket{0}\otimes \ket{\tilde{m}_+}$, $\ket{M_-}=\ket{0}\otimes \ket{\tilde{m}_-}$.
Then, we can write the cavity-molecule Hamiltonian in the following form: 
\begin{align}\label{eq:mono1} 
H& =\begin{pmatrix} 
\omega_{1} + \Delta & {g_{+}}  & {g_{-}} \\
{g_{+}}  & E_{+} & 0 \\
{g_{-}}  & 0 & E_{-}
\end{pmatrix},
\end{align}
where $\Delta = \omega_{c}-\omega_{1}$ is the detuning between the cavity mode and the $QD_1$. Besides, ${g_{+}}$ and ${g_{-}}$ are related to the original light-matter interaction constants ${g_{1}}$ and ${g_{2}}$ by:
\begin{align}
g_{+}&=g_{1}\cos\theta_m + g_{2}\sin\theta_m; \label{eq:gp} \\
g_{-} &= -g_{1}\sin\theta_m + g_{2}\cos\theta_m. \label{eq:gm}
\end{align}
Here, the  angle $\theta_m$ is determined by
\begin{align}\label{eq:thpp}
\tan 2\theta_m = \frac{2T}{\omega_1 -\omega_2}.
\end{align}

\subsubsection{ Polaritonic Basis Picture}
Similarly, we can build up another basis that takes into account a dressed state of a photon interacting with a single quantum dot. In this case, we consider the bound state between light and the QD$_{1}$, i.e., the Hamiltonian for the photon-QD$_{1}$ subsystem is  
\begin{equation}\label{eq:Hami3}
H_{Cav-QD_1}=\left (\begin{array}{cc}
\omega_{c}  & g_{1}  \\ 
g_{1} &  \omega_{1}\\
\end{array}\right),
\end{equation}\\
that has the following dressed eigenstates,
\begin{align}
\ket{\tilde{p}_{+}}&=\cos\theta_{p}\ket{0,X}+\sin\theta_{p}\ket{1,G},\\  \ket{\tilde{p}_{-}}&=-\sin\theta_{p}\ket{0,X}+\cos\theta_{p}\ket{1,G},
\end{align}
and we define a polaritonic ground state like $\ket{\tilde{p}_{0}}=\ket{0,G}$.   
The corresponding eigenvalues are:
\begin{align}
\label{eq:avalll}
\textrm{E}^{'}_{\pm}&= \dfrac{1}{2} (2\omega_{1}+\Delta) \\ \nonumber
&\pm \dfrac{1}{2} \sqrt{(2\omega_1+\Delta)^2-4 \omega_1 (\omega_1 + \Delta)+4g_{1}^2},
\end{align}
also, with this polaritonic basis, we propose a new basis for the whole system, that corresponds to: $\ket{P_0}=\ket{\tilde{p}_0}\otimes\ket{X}$, $\ket{P_+}= \ket{\tilde{p}_+}\otimes\ket{G}$, $\ket{P_-}=\ \ket{\tilde{p}_-}\otimes\ket{G}$.
Then, we can write the polariton-QD$_{2}$ Hamiltonian in the following form: 
\begin{align}\label{eq:pol} 
H &= \begin{pmatrix} 
E_{+}^\prime & 0 & g_{1}^\prime\\
0 & E_{-}^\prime & g_{2}^\prime\\
g_{1}^\prime & g_{2}^\prime & \omega_{2}
\end{pmatrix},
\end{align}
here $g_{1}^\prime$ and $g_{2}^\prime$ are related to the original light-matter interaction constants ${g_{2}}$ and tunnelling $T$ by:
\begin{align}\label{eq:g22}
g_{1}^\prime&=g_{2}\cos\theta_{p} + T \sin\theta_{p};\\
g_{2}^\prime &= -g_{2}\sin\theta_{p} + T\cos\theta_{p},
\end{align}
where the angle $\theta_{p}$ is determined by \begin{align}\label{eq:thpp2}
\tan 2\theta_{p} = \frac{2g_1}{\Delta}.
\end{align}
In what follows we refer to the eigenenergies of the eigenstates $\ket{\lambda_+}$, $\ket{\lambda_0}$, $\ket{\lambda_-}$ of the Hamiltonian (\ref{eq:Hami2}), (\ref{eq:mono1}) or (\ref{eq:pol}) as
$\lambda_+$, $\lambda_0$ and $ \lambda_-$, respectively.

\begin{figure}[thb]
\centering
\includegraphics[width=0.5\textwidth]{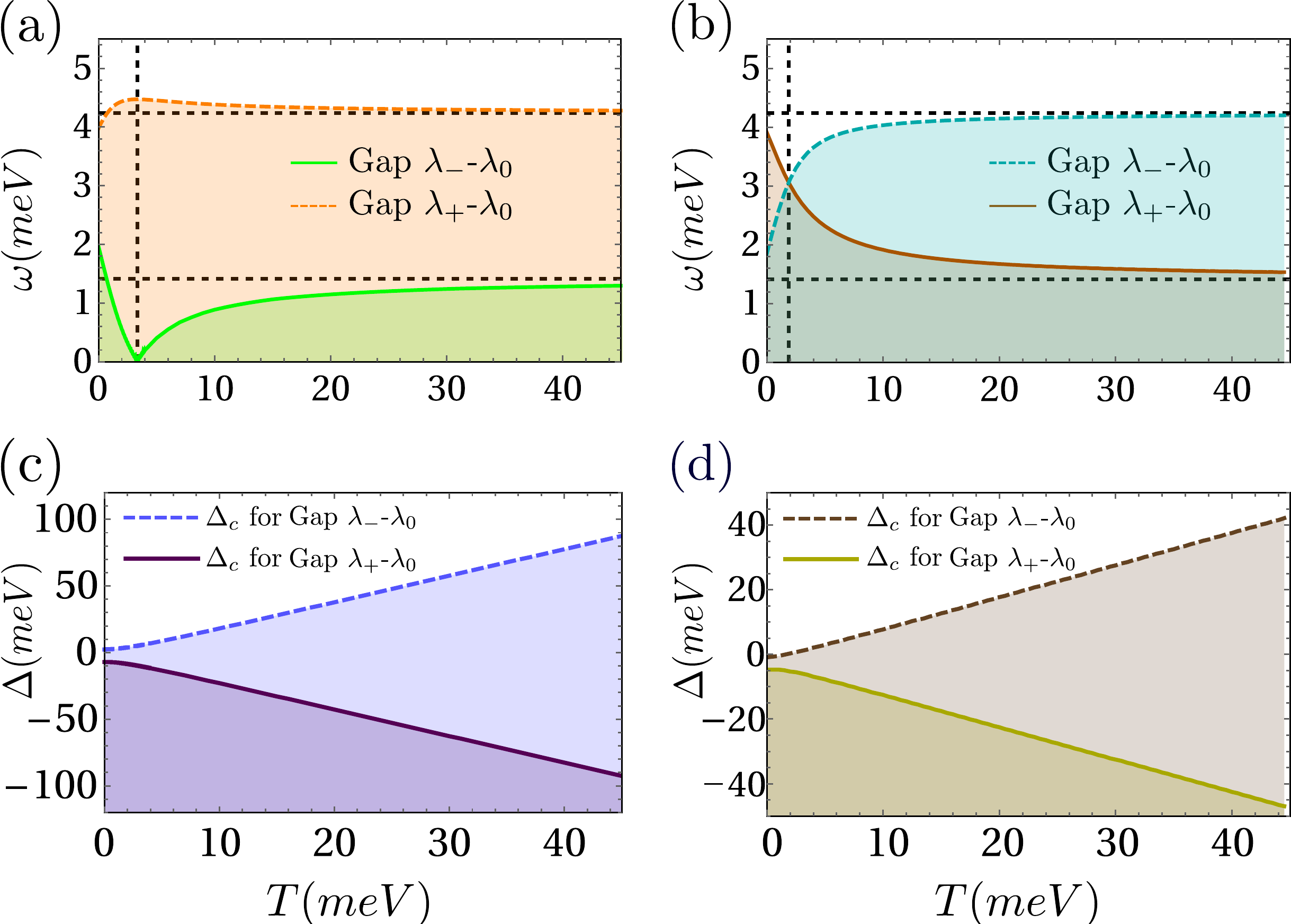}
\vspace{-0.4cm}
\caption{(a)-(b) The bandgaps: $\lambda_- -\lambda_0$ (green)-(cyan) and $\lambda_+ -\lambda_0$ (orange)-(brown), as a function of the tunnelling strength $T$. The parameters used are: $\omega_{1}=1005\,$meV, $\omega_{2}=1000\,$meV in all cases, $g_{1}=2\,$meV, $g_{2}=1\,$meV in (a)-(c) and $g_{1}=1\,$meV, $g_{2}=2\,$meV in (b)-(c). (c)-(d) show bandgaps positions $\Delta_c$ as function of $T$. Vertical dashed lines identify critical bandgap conditions $T_c=3.\overline{3}\,$meV ---(a) case 1--- and $T_c=1.875\,$meV ---(b) case 2---. Horizontal dashed lines show the bandgaps asymptotic behaviour.}
\label{fig:gaps}
\end{figure}

\section{Results and discussion}

\begin{figure*}
\centering
\includegraphics[width=1.0\textwidth]{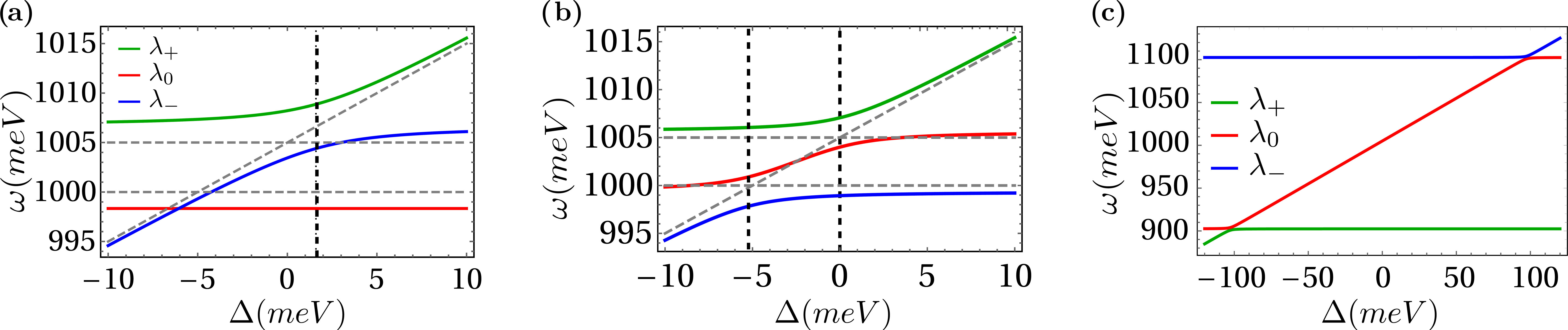}
\vspace{-0.4cm}
\caption{Energy eigenvalues for the three different regimes. (a) Case 1: Maximal first bandgap and complete suppression of the second bandgap with:  $g_{1}=2\,$meV, $g_{2}=1\,$meV, $T=3.\overline{3}\,$meV. (b) Case 2: Simultaneous maximum gap condition with: $g_{1}=1\,$meV, $g_{2}=2\,$meV, $T=1.875\,$meV. (c) Case 3: Strong molecular coupling with: $g_{1}=1\,$meV, $g_{2}=2\,$meV, $T=100\,$meV. In all cases $\omega_{1}=1005\,$meV and $\omega_{2}=1000\,$meV. Grey dashed lines show the bare energies.}
\label{fig:unov4}
\end{figure*}

Because the emission of light is related to the gap structure in the energy band diagram, we plot the two bandgaps dependence with tunnelling strength for two different operational regimes. In the first one, Fig.~\ref{fig:gaps} (a), $g_1>g_2$, $\omega_1-\omega_2>0$,  and $g_-=0$, condition which yields $T=(\omega_1-\omega_2)g_1g_2/(g_1^2-g_2^2)$. This specific feature occurs when the first bandgap is maximal and the second bandgap is completely suppressed (vertical dashed line). In the second regime, Fig.~\ref{fig:gaps} (b),  $g_1<g_2$, $\omega_1-\omega_2>0$, and $g_+=g_-$, condition which yields $T=(\omega_1-\omega_2)(g_2^2-g_1^2)/4g_1g_2$. Here both bandgaps are the same and simultaneously maximal (vertical dashed line).
Note that, in the bandgap's figures, the values of   $\lambda_- -\lambda_0$ and $\lambda_+ -\lambda_0$ have the same asymptotic limit as $T \to \infty$  (horizontal dashed lines). These quantities  correspond to $\sqrt{2}(g_1+g_2)$ and $\sqrt{2}(g_1-g_2)$. Particularly, for both set of parameters used, they take the numerical values $3 \sqrt{2}\,$meV, and $\sqrt{2}\,$meV. In the limit of large molecular coupling, two  eigenvalues have a linear behaviour with $\pm T$, and the other one is linear in $\Delta$. Besides, the critical detuning $\Delta_c$ also follows the same linear behaviour. This fact entails the limiting previously explained bandgaps conditions. 
Figs.~\ref{fig:gaps} (c)--(d) show bandgaps position $\Delta_c$ as a function of $T$; despite the apparent similarity of these two plots, it is possible to observe the detuning shift for each regime. In this scenario, we consider three particular situations. Case 1: maximal first bandgap and complete suppression of the second bandgap; case 2: simultaneous maximal gap condition; case 3: strong  molecular coupling.

\subsection{Case 1: Maximal first bandgap and complete suppression of the second bandgap}

Making $g_{-}$ equal to zero in expression (\ref{eq:gm}), what suppresses the first bandgap, we can solve for $\theta$ in the interval $[0,\pi/2]$. In this case, the cavity mode decouples from the lower molecular state yielding $\ket{M_{-}}$. The Fig.~\ref{fig:unov4}~(a) shows the complete energy eigenvalues for the system: $\lambda_{-}$, $\lambda_{0}$, $\lambda_{+}$. This plot was made with the following parameters: $\omega_{1}=1005\,$meV,                   $\omega_{2}=1000\,$meV, $g_{1}=2\,$meV, $g_{2}=1\,$meV, $T=3.\overline{3}\,$meV. The first energy bandgap is suppressed for $\Delta=-6.1\,$meV and this condition maximises the second bandgap for $\Delta=1.6\,$meV; the corresponding bandgaps are $\omega=0\,$meV and $\omega=4.47\,$meV, respectively.

\begin{figure}[H]
\centering
\includegraphics[width=0.45\textwidth]{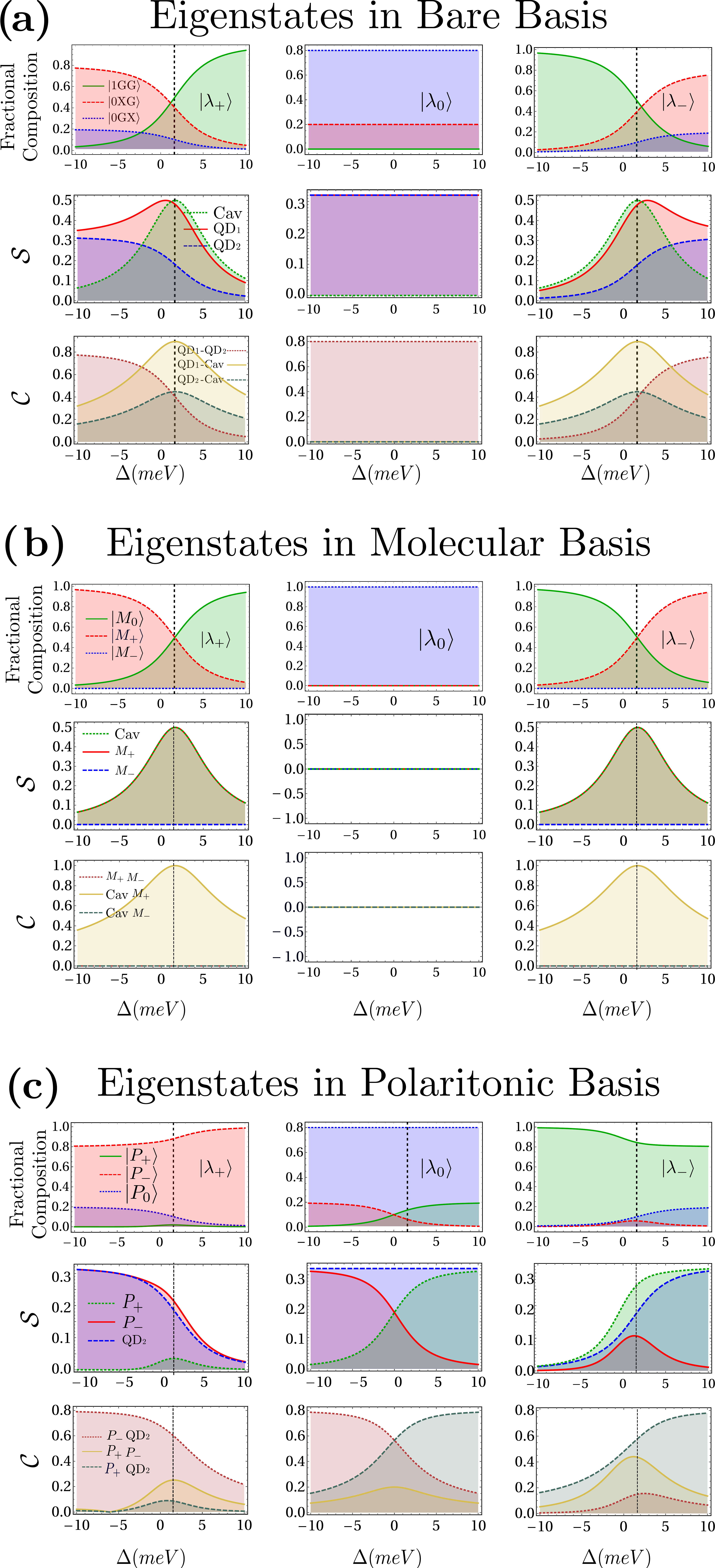}
\caption{Case 1: Maximal first bandgap and complete suppression of thy second bandgap. The set of parameters is: $\omega_{1}=1005\,$meV, $\omega_{2}=1000\,$meV, $g_{1}=2\,$meV, $g_{2}=1\,$meV, $T=3.\overline{3}\,$meV. (a) Fractional Composition, Linear Entropy and Concurrence for the Eigenstates in Bare Basis. (b) and (c) for the Eigenstates in Molecular Basis and Eigenstates in Polaritonic Basis, respectively.}
\label{fig:unov5}
\end{figure}

In Fig.~\ref{fig:unov5} we depict three panels for the Case 1: a) Eigenstates in Bare Basis, (b) Eigenstates in Molecular Basis, and (c) Eigenstates in  Polaritonic Basis. Each one contains the fractional composition, the linear entropy calculated by $S_L = 1 − Tr(\rho^2)$~\cite{entropy}  and the concurrence calculated by $C(\rho)= max \{0,\lambda_1 -\lambda_2 -\lambda_3-\lambda_4\}$, where the $ \lambda_i$'s are the square roots of the eigenvalues of $ \rho\tilde\rho$ in descending order. Here $ \tilde\rho$ is the result of applying the spin-flip operation to $\rho$: $\tilde\rho=(\sigma_y\otimes\sigma_y)\rho^*(\sigma_y\otimes\sigma_y)$, and the complex conjugation is again taken in the $\sigma_z$ basis  ~\cite{concurrence}.
Panel (a) shows the fractional composition as a function of detuning $\Delta$; it indicates that this description fails to provide a useful quasiparticle scheme to understand the whole physical mechanisms that are involved in this regime. Moreover, it shows relevant contributions of each component (bare states) for different values of $\Delta$. In particular, we point out the value of $\Delta=1.6\,$meV, where the anticrossing appears. Linear entropy and concurrence confirm that no bare state is enough to comprehend the main features of entanglement of the system, they only show entanglement between partial subsystems; then, a global quasiparticle cannot be appreciated.

The Molecular Basis is used to plot the Hamiltonian eigenstates in panel (b). Once again, contrary to intuition, this basis fails to give a whole quasiparticle scheme. In this case, only the vector $\ket{M_{-}}$ describes exactly an Hamiltonian eigenstate $\ket{\lambda_{0}}$; but the other two, $\ket{\lambda_{+}}$ and $\ket{\lambda_{-}}$, are mixed compositions of the vectors $\ket{M_{0}}$ and $\ket{M_{+}}$. Because the state $\ket{M_{-}}$ is decoupled from the other two states, $\ket{M_{0}}$ and $\ket{M_{+}}$, the linear entropy and the concurrence are the same, but it takes only a maximum value in the anticrossing condition.

Finally, panel (c) displays a description in the Polaritonic Basis. In this case, the fractional composition shows that each vector of this basis identifies by itself the quantum correlations of the Hamiltonian eigenstates. The Hamiltonian is almost diagonal in this polaritonic basis; therefore it gives a useful quasiparticle scheme to interpret the underlying physical processes.

\subsection{Case 2: Simultaneous maximal gap condition}

\begin{figure}[H]
\centering
\includegraphics[width=0.45\textwidth]{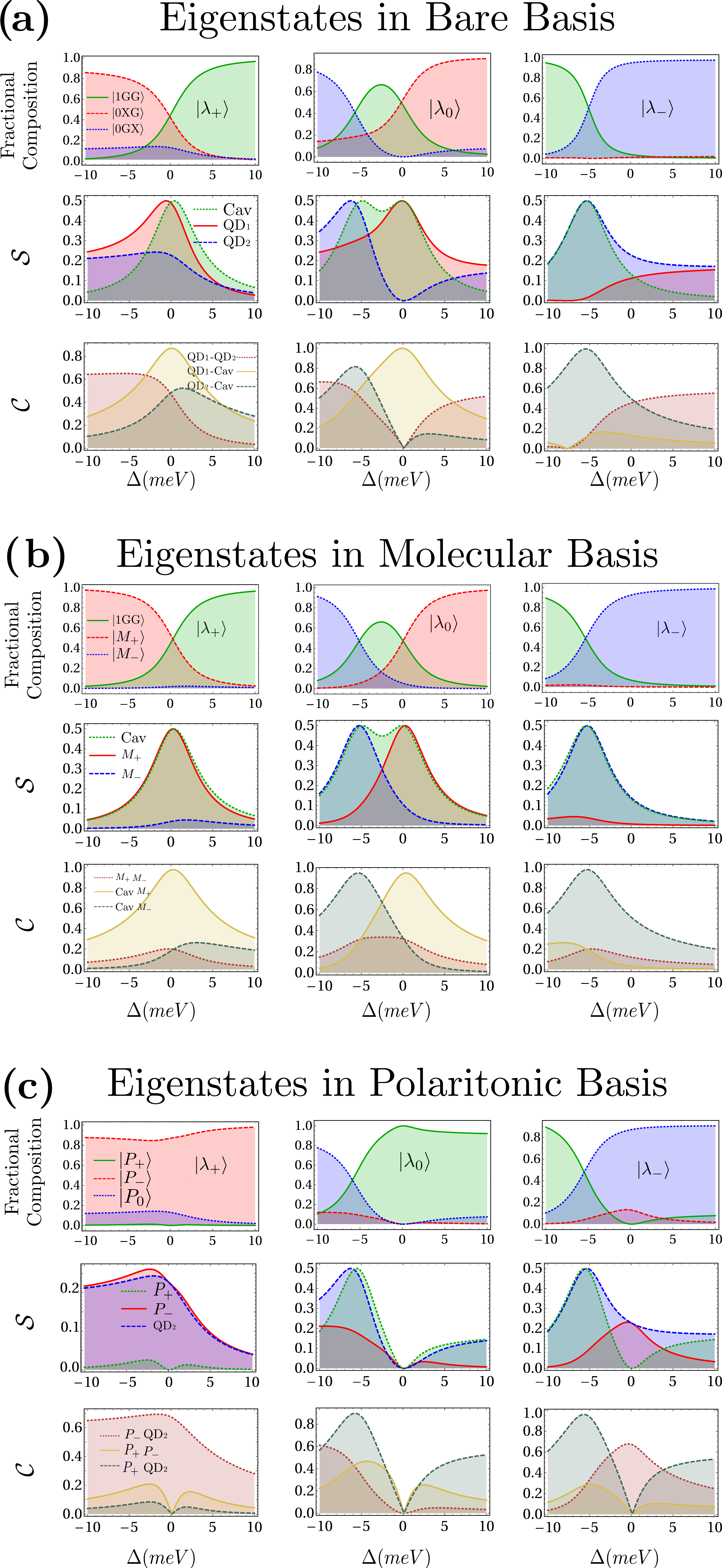}
\caption{Case 2: Simultaneous maximum gap condition. The set of parameters are:  $\omega_{1}=1005\,$meV, $\omega_{2}=1000\,$meV, $g_{1}=1\,$meV, $g_{2}=2\,$meV, $T=1.875\,$meV. (a) Fractional Composition, Linear Entropy and concurrence for the Eigenstates in Bare Basis. (b) and (c) for the Eigenstates in Molecular Basis and Eigenstates in Polaritonic Basis, respectively.}
\label{fig:caseuno}
\end{figure}

In Case 2, we consider $g_{+}=g_{-}$ in the expression (\ref{eq:mono1}). This condition provides the simultaneous maximisation of energy bandgaps in both anticrossings as it is shown in Fig.~\ref{fig:unov4} (b).  This anticrossings appear in values of $\Delta=-5.1\,$meV and $\Delta=0.1\,$meV (vertical dashed lines) and the corresponding bandgap in each $\Delta$ is $\omega=3.05\,$meV. In this case, we can see again that the bare basis cannot catch an effective description in terms of quasiparticles for the whole system. This is because, in this basis, the entanglement properties between the component states are maximal at different values of $\Delta$ as it is shown by the linear entropy and concurrence in fig.~\ref{fig:caseuno}~(a). The last argument is again valid for the case of molecular basis, Fig.~\ref{fig:caseuno}~(b), despite that the number of maxima of entanglement is diminished, they are still  present. Therefore, this representation is not convenient. Finally, the polaritonic representation is, again, able for tackling the main features of this regime. The fractional composition shows that is possible to identify many regions where the polariton representation captures the essential structure of the Hamiltonian eigenstates. For instance, the linear entropy for the eigenstates $\ket{\lambda_+}$, $\ket{\lambda_0}$, and $\ket{\lambda_-}$ shows that, at resonance $\Delta=0\,$meV, they can be well described by $\ket{P_-}$, $\ket{P_+}$, and $\ket{P_0}$, respectively. A good criterion to determine if the basis match well with the Hamiltonian eigenstates is to get shallow values of linear entropy, $\mathcal{S}$, 

\begin{figure}[H]
\centering
\includegraphics[width=0.45\textwidth]{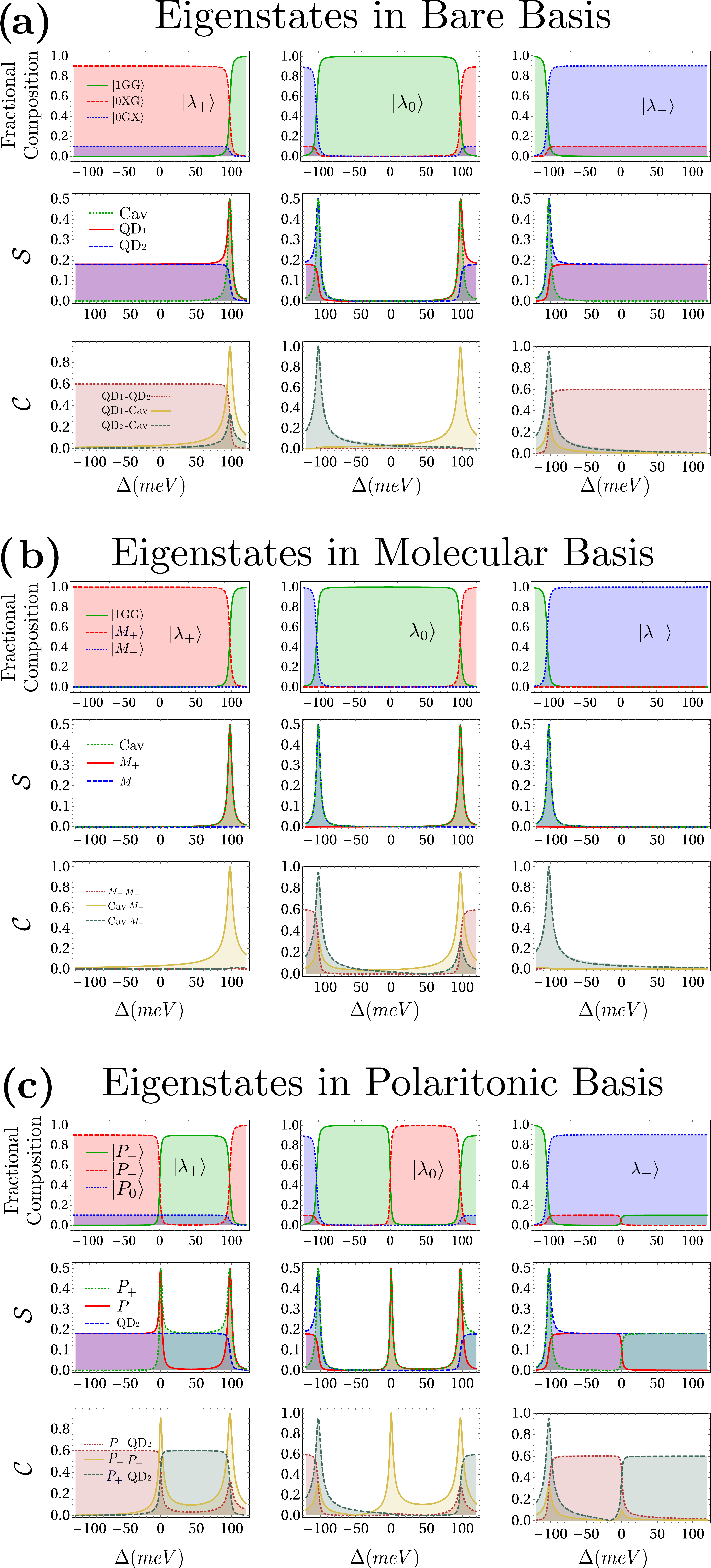}
\caption{Case 3:   Strong molecular coupling.  The set of parameters are: $\omega_{1}=1005\,$meV, $\omega_{2}=1000\,$meV, $g_{1}=1\,$meV, $g_{2}=2\,$meV, $T=100\,$meV. (a) Fractional Composition, Linear Entropy and Concurrence for the Eigenstates in Bare Basis. (b) and (c) for the Eigenstates in Molecular Basis and Eigenstates in Polaritonic Basis, respectively.}
\label{fig:unov3}
\end{figure}
\noindent which is calculated, for each quasiparticle, tracing over the two remaining subsystems. 

\subsection{Case 3: Strong Molecular Coupling}

As in the two cases above, we consider another situation of interest: Strong molecular coupling where $T\gg g_1,g_2$. In this regime the Hamiltonan, in the molecular basis, reads
\begin{align}\label{eq:HamTM} 
H& =\begin{pmatrix} 
\Delta & {g_{+}}  & {g_{-}} \\
{g_{+}}  & T & 0 \\
{g_{-}}  & 0 & -T
\end{pmatrix},
\end{align}
here, the molecular eigenstates have a linear behaviour with $\pm T$.

Two anticrossings are shown in Fig.~\ref{fig:unov4} (c) for $T=100\,$meV. They appear in $\Delta = -102.51\,$meV and  $\Delta = 97.53\,$meV.  Fig.~\ref{fig:unov3}~(a)-(b)-(c) show that all basis are relatively good for grasping the essential features of the system in terms of a quasiparticle behaviour. The molecular basis, Fig.~\ref{fig:unov3}~(b), does catch the desired behaviour in an optimal way across the detuning range. The fractional composition shows that the system is found prepared in a pure state of each molecular mode. Moreover, the system gets suddenly entangled, i.e., each molecular vector and the photon shows a sharp peak of concurrence and linear entropy. Despite, in this case, the molecular basis seems to be a more suitable basis for a successful description. Bare basis and polariton basis,  Fig.~\ref{fig:unov3}~(a)-(c), remain to be able to capture almost all the essential features of the system.

\section{Conclusions}

A comparison of three effective pictures of quasiparticles was performed to represent the Hamiltonian eigenstates for a two quantum dot-microcavity system. As a general conclusion, we demonstrate that the polariton basis, i.e., dressed states of photon and exciton, can depict in different regimes the physics contained in the Hamiltonian model we are dealing with. This theoretical tool can be useful to seek physical regimes where collective quasiparticle modes are required and it gives experimental insights to find them.

\section*{Acknowledgments}

F.G and B.A.R. acknowledge financial support from Codi-UdeA trough ``Proyecto de Sostenibilidad del Grupo de Física Atómica". J.P.R.C. and H.V-P acknowledge financial support from UN--DIEB project ``Control dinámico de la emisión en sistemas de Qubits acoplados con cavidades no-estaciona\-rias" HERMES 41611; and from COLCIENCIAS under the project ``Emi\-sión en sistemas de Qubits Superconductores acoplados a la radiación", Código 110171249692, CT 293-2016, HERMES 31361.  J.P.R.C. gratefully acknowledges financial support from the ``Beca de Doctorados Nacionales de COLCIENCIAS 785".

\bibliographystyle{elsarticle-num} 
\bibliography{Ref.bib}

\end{document}